\documentclass[sigplan,10pt]{acmart}
\usepackage{graphicx}     
\usepackage{enumitem}
\usepackage{multirow}

\usepackage{xcolor}      
\usepackage{listings}    
\usepackage{etoolbox}    
\usepackage{tabularx}
\usepackage{array} 
\usepackage{ragged2e}   
\usepackage{booktabs}
\usepackage{hyperref}
\usepackage{amsmath}
\usepackage{siunitx}
\usepackage{soul}
\usepackage[linesnumbered,ruled,vlined]{algorithm2e}

\usepackage{tikz}

\newcolumntype{C}[1]{>{\centering\arraybackslash}m{#1}}
\newcolumntype{Y}{>{\raggedright\arraybackslash}X}

\usetikzlibrary{arrows.meta,calc,positioning,fit,decorations.pathmorphing}

\definecolor{deepgreen}{rgb}{0,0.5,0}
\definecolor{deepblue}{rgb}{0,0,0.4}
\definecolor{deepred}{rgb}{0.6,0,0}
\lstdefinestyle{pythoninline}{
  language=Python,
  basicstyle=\ttfamily\color{red!80!black},
  keywordstyle=\color{purple!80!black},
  stringstyle=\color{orange!80!black},
  commentstyle=\color{gray}\itshape,
  breaklines=true,
  showstringspaces=false
}
\newcolumntype{C}[1]{>{\centering\arraybackslash}m{#1}}
\newcolumntype{Y}{>{\RaggedRight\arraybackslash}X}

\acmConference[PPoPP'26]{ACM SIGPLAN Annual Symposium on Principles and Practice of Parallel Programming}{January 31--February 04, 2026}{Sydney, Australia}
\acmYear{2026}
\acmISBN{} 
\acmDOI{} 
\startPage{1}

\setcopyright{none}

\usepackage{booktabs}   
\usepackage{subcaption} 

\begin{document}

\title{PolyKAN: Efficient Fused GPU Operators for Polynomial Kolmogorov–Arnold Network Variants}

\author{Mingkun Yu}
\affiliation{%
  \institution{Sun Yat-sen University}
  \city{Guangzhou}
  \country{China}
}

\author{Heming Zhong}
\affiliation{%
 \institution{Sun Yat-sen University}
 \city{Guangzhou}
 \country{China}}

\author{Dan Huang}
\affiliation{%
  \institution{Sun Yat-sen University}
  \city{Guangzhou}
  \country{China}}

\author{Yutong Lu}
\affiliation{%
  \institution{Sun Yat-sen University}
  \city{Guangzhou}
  \country{China}}

\author{Jiazhi Jiang}
\affiliation{%
  \institution{Sun Yat-sen University}
  \city{Guangzhou}
  \country{China}}

\fancyhead{}  
\renewcommand\footnotetextcopyrightpermission[1]{} 

\begin{abstract}
Kolmogorov–Arnold Networks (KANs) promise higher expressive capability and stronger interpretability than Multi-Layer Perceptron, 
particularly in the domain of AI for Science. However, practical adoption has been hindered by low GPU utilization of 
existing parallel implementations. To address this challenge, we present a GPU-accelerated operator library, named PolyKAN 
which is the first general open-source implementation of KAN and its variants. 
PolyKAN fuses the forward and backward passes of polynomial KAN layers into a concise set of optimized CUDA kernels. 
Four orthogonal techniques underpin the design: (i) \emph{lookup-table} with linear interpolation that replaces runtime expensive math-library functions; (ii) \emph{2D tiling} to expose thread-level parallelism with preserving memory locality; (iii) a \emph{two-stage reduction} scheme converting scattered atomic updates into a single controllable merge step; and (iv) \emph{coefficient-layout reordering} yielding unit-stride reads under the tiled schedule. 
Using a KAN variant, Chebyshev KAN, as a case-study, PolyKAN delivers $1.2$--$10\times$ faster inference and $1.4$--$12\times$ faster training than a Triton + cuBLAS baseline, with identical accuracy on speech, audio-enhancement, and tabular-regression workloads on both highend GPU and consumer-grade GPU. 
\end{abstract}


\maketitle

\begin{figure}[htbp]
  \centering
  \includegraphics[width=\linewidth]{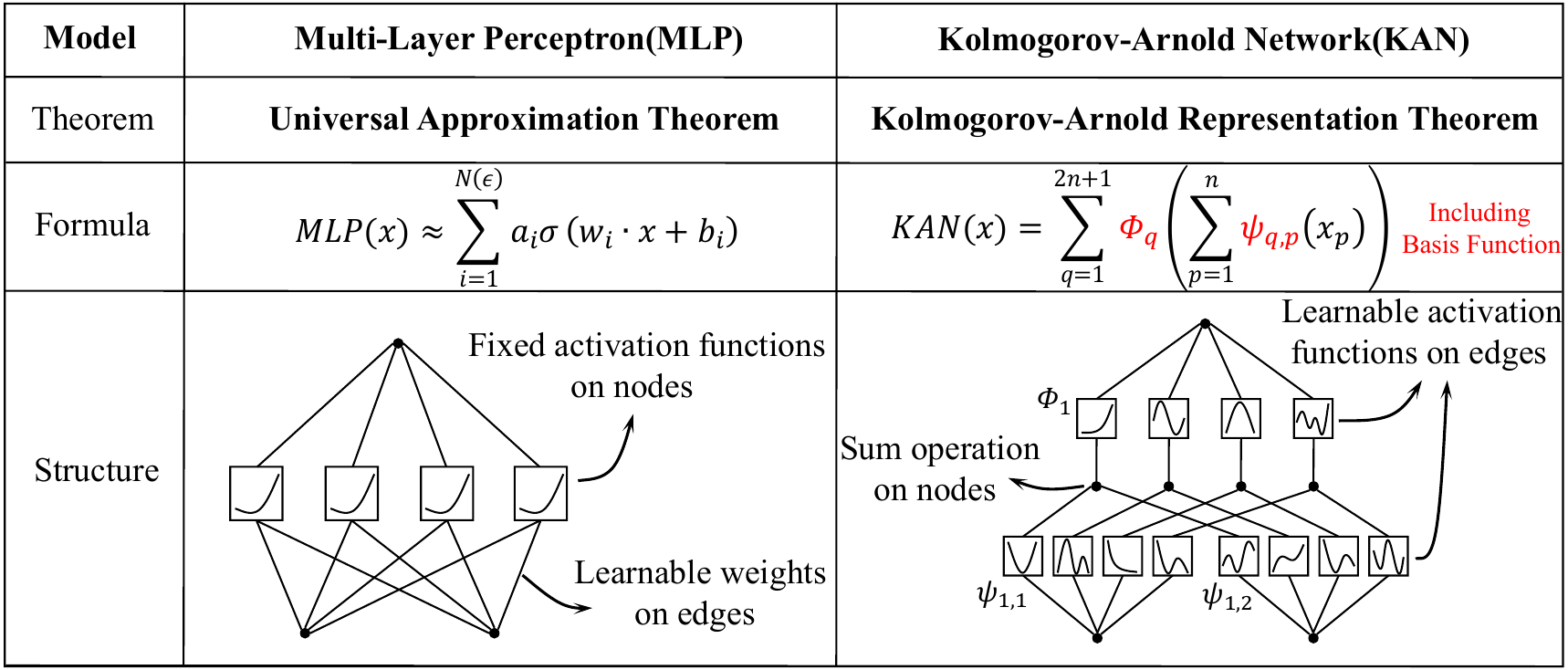}
  \caption{Architectural and theoretical comparison between traditional multi-layer perceptron (MLP) and Kolmogorov-Arnold Network (KAN).}
  \label{fig:mlp_vs_kan}
\end{figure}

\section{Introduction}
Deep learning (DL) has achieved remarkable progress across domains such as computer vision, natural language processing, 
and scientific computing \cite{lecun2015deep}. Multilayer perceptrons (MLPs) \cite{Rumelhart_Hinton_Williams_1986} are foundational 
building blocks of deep learning, yet their inherently opaque nature raises concerns about transparency and interpretability 
\cite{rudin2019stop,cybenko1989approximation}. To pursue more accuracy and higher interpretability, researchers are compelled to 
explore novel model architectures and activation mechanisms of deep learning. As illustrated in Figure \ref{fig:mlp_vs_kan}, 
traditional MLP employs fixed non-linear activation functions such as ReLU \cite{nair2010rectified}, Sigmoid \cite{Rumelhart_Hinton_Williams_1986} 
and Tanh \cite{glorot2010understanding}. In contrast, Kolmogorov-Arnold Network (KAN)~\cite{liu2024kan} replaces the fixed activation functions 
with a linear combination of a set of polynomial basis functions, based on Kolmogorov-Arnold representation theorem~\cite{Kolmogorov}. 
The specific pattern of linear combination is determined by a set of learnable coefficients. Therefore, the process of mapping the input 
to the output through the nonlinear activation function is transparent.

KANs offer improved memory capacity, interpretability, and accuracy compared to traditional MLPs \cite{yu2024kan}. 
Therefore, KANs have been successfully extended to reconstruct various neural network modules, including convolutional \cite{bodner2024convolutional}, 
graph architectures \cite{zhang2024graphkan}, even Transformer \cite{yang2024kolmogorov} and large language models \cite{GANESH2024KANGPT}. 
Especially in the domain of AI for Computational Science and Engineering such as partial differential equation \cite{Evans2010PDE}, 
KAN has shown much better performance than MLP due to its characteristics \cite{wang2025physics, yang2025multi, guo2024physics}. 
To adapt to different tasks, KANs can further enhance the capability through adjusting basis functions and parameterization configurations. This prompts a wide spectrum of KAN variants based on Fourier \cite{xu2024fourierkan}, Chebyshev \cite{ss2024chebyshev}, Legendre \cite{torchkan}, and other basis functions.

Although KAN variants possess these unique advantages, they typically suffer from $10\times$ slower runtimes than MLPs with comparable model 
and parameter sizes \cite{liu2024kan}. This inefficiency stems from: (i) the use of parameterized univariate functions as activation function substantially 
increases computational overhead, (ii) KAN basis-expansion primitives use naive loop-based implementation, limited optimization for parallelism strategies, such as kernel fusion, and (iii) irregular memory access limits GPU concurrency. Convolutional and GEMM benefit from deeply optimized libraries (e.g., cuDNN \cite{chetlur2014cudnn}, cuBLAS \cite{nvidia_cublas_2025}). In contrast, polynomial basis expansion, which uses a linear combination of basis polynomial functions to represent
complex functions, still lacks a high-performance kernel library, resulting in a major bottleneck for practical KAN deployment.

To address this issue, we propose a systematic approach of GPU parallel
optimization for KAN and its variants, exemplified by Chebyshev KAN (ChebyKAN). Our approach combines lookup-table (LUT) interpolation to alleviate the high cost of polynomial basis expansions, 2D tiling over inputs and outputs to improve spatiotemporal locality, two-stage reduction to mitigate atomic contention, and coefficient-layout reordering for coalesced access. This approach offers a reusable operator interface for seamless integration to prevalent deep learning frameworks, such as PyTorch. Our main contributions are summarized as follows:

\begin{enumerate}[label=\arabic*),leftmargin=*,align=parleft]
    \item \textbf{Systematic analysis of the core bottlenecks in KAN-type networks.} We identify the issues of ``multi-step dependency'' and ``complex function calls'' for high-order polynomials under GPU parallelism. Furthermore, we analyze how traditional, operator-by-operator concatenation fails to fully exploit GPU potential from both computational and memory-access perspectives.
    
    \item \textbf{A general, extensible fused-kernel design paradigm.} We propose a general fused-kernel paradigm that integrates forward/backward computations with LUT-based evaluation, 2D tiling, two-stage reduction, and coefficient layout reordering, significantly reducing kernel launch overhead and atomic conflicts. This results in $1.3$--$2.2\times$ speedup and $1.3$--$4\times$ throughput improvement on end-to-end tasks.
    
    \item \textbf{Generalization across KAN variants.} We analyze the computational characteristics of the KAN variant and demonstrate the generalization of our proposed method. The proposed fused-kernel design is independent of basis function selection, supporting KAN variants based on Chebyshev, Legendre, Fourier, etc. Additionally, the design can function as a plug-in component, enabling its seamless integration into complex model architectures such as Convolutional Networks, Graph Neural Networks, and Transformers.
    
    \item \textbf{A reusable operator optimization library, PolyKAN, for polynomial/kernel approximation networks.} We implement and evaluate an open-source library, \emph{PolyKAN}, delivering substantial training and inference speedups without accuracy loss, and providing Python APIs for better usability in domain of AI for Science. To the best of our knowledge, this is the first open-source, general GPU operator library for the polynomial-based KAN variants.
\end{enumerate}

\noindent The rest of the paper overviews KAN and polynomial expansions (\S 2), analyzes bottlenecks and motivation (\S 3), details our optimizations (\S 4), and presents experiments (\S 5) before concluding (\S 6).

\section{Background and Related Work}
\subsection{Overview of KAN}

As an alternative to traditional MLPs, KAN introduces a novel architecture that replaces fixed node-wise activations with learnable edge-wise univariate functions, aiming to improve both expressive efficiency and interpretability. This design is grounded in the idea that complex multivariate continuous functions can be decomposed into compositions of simpler univariate ones. The theoretical underpinning of this decomposition is the Kolmogorov-Arnold Theorem, which we briefly review below.

\begin{theorem}[Kolmogorov-Arnold Theorem \cite{Kolmogorov}]
Let \(f: [0,1]^n \allowbreak \to \mathbb{R}\) be an arbitrary continuous function. Then there exist continuous single-variable functions
\[
\phi_{0},\phi_{1}, \dots, \phi_{2n}
\ \text{and}\ 
\psi_{q,1}, \psi_{q,2}, \dots, \psi_{q,n} \quad (\text{for } 1 \leq q \leq  2n+1)
\]
such that, for all \(x = (x_1, x_2, \ldots, x_n) \in [0,1]^n\),
\[
f(x) \;=\; \sum_{q=1}^{2n+1} \phi_q\!\biggl(\,\sum_{p=1}^n \psi_{q,p}(x_p)\biggr).
\]
\end{theorem}

\begin{figure}[htbp]
  \centering
  \includegraphics[width=0.9\linewidth]{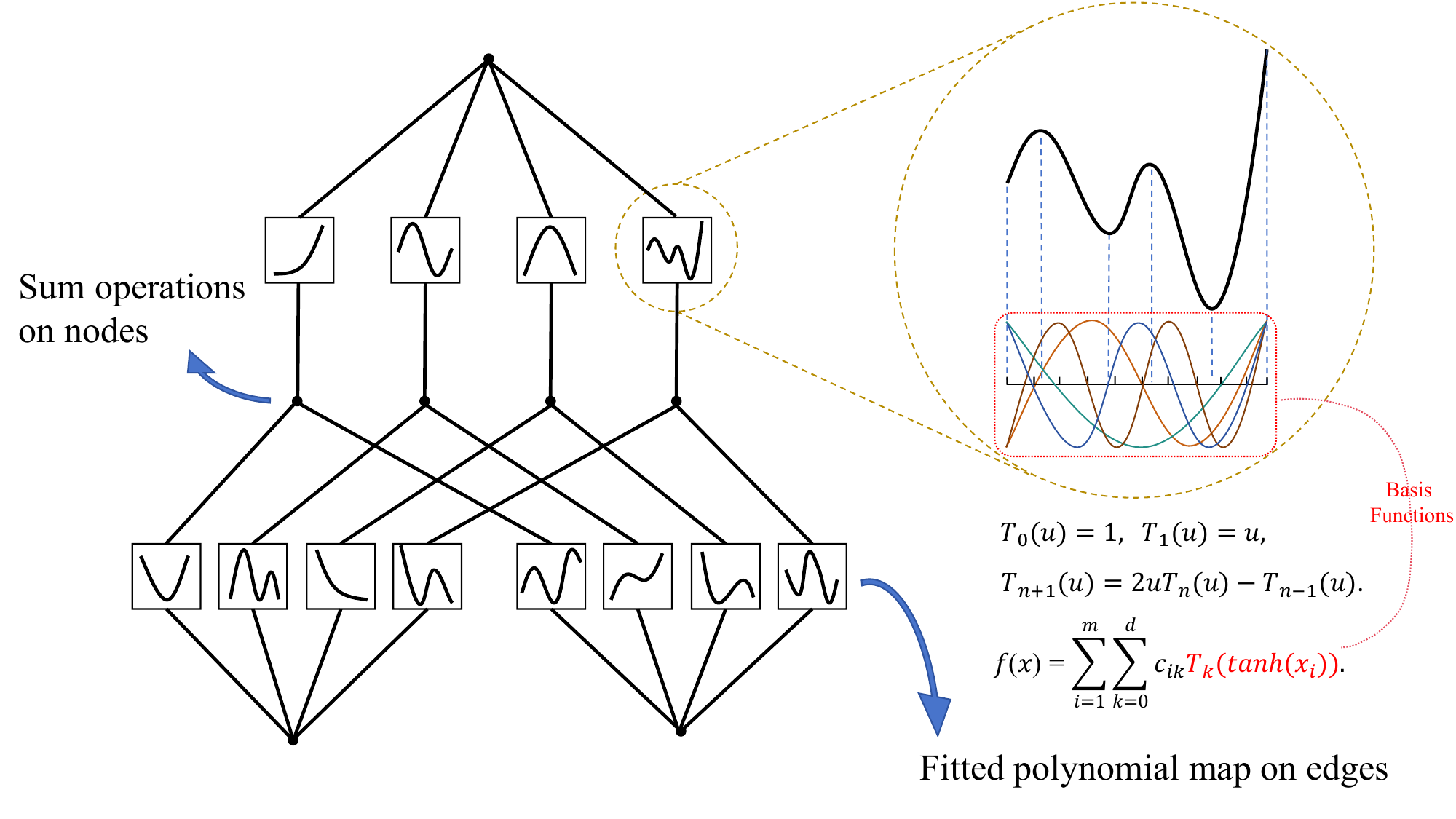}
  \caption{The structure of the Kolmogorov-Arnold network.}
  \label{fig_2}
\end{figure}

Based on the Kolmogorov-Arnold theorem, KAN adopts a network structure that can be broadly described as follows: 
for an arbitrary input vector $\mathbf{x} \in \mathbb{R}^n$, each component $x_p$ is first mapped by a univariate function 
$\psi_{q,p}$, where $p$ indexes the dimensions of $x$ and $q$ indexes channels. The results of these mappings across all input 
dimensions are then summed, and the summed value is passed through another univariate function $\phi_q$. Finally, the scalar outputs 
from all $q$ paths are added together to produce the final output. The structure of the Kolmogorov-Arnold network is shown in Figure \ref{fig_2}.

\subsection{KAN variant: Chebyshev KAN}
\label{fomula}
Chebyshev KAN \cite{ss2024chebyshev} is a variant of KAN. The univariate basis functions are implemented with Chebyshev polynomials, replacing the conventional B‑spline basis or other activation functions. The \textbf{\textit{first computation strategy}} for the Chebyshev polynomial expansion is defined as follows:
\begin{equation}
T_{n}(x) = \cos\bigl(n\arccos x\bigr), \qquad x \in [-1,1],\  n \in \mathbb{N}.
\label{th:chebyshev}
\end{equation}
where $n$ denotes the order of the Chebyshev polynomial, which governs both its shape and approximation 
capacity. When a target function exhibits greater complexity, $n$ typically resorts to polynomials of higher order. 
Since GPUs have high throughput on basic operators such as vectorized addition, subtraction, multiplication and 
division, the overhead of calling high-level functions such as $\cos(\cdot)$ or $\sin(\cdot)$ multiple 
times is much higher than addition and multiplication. Therefore, exploiting the trigonometric identity
$\cos\bigl((n+1)\theta\bigr)=2\cos\theta\cos(n\theta)-\cos\bigl((n-1)\theta\bigr),$ investigators 
establish the \textbf{\textit{second computation strategy}} of Chebyshev polynomial expansion 
by the following recurrence formula:
\begin{equation}
T_{0}(x) = 1,\;\;\; T_{1}(x) = x, \;\;\; T_{n+1}(x) = 2xT_{n}(x)-T_{n-1}(x).
\label{re:chebyshev}
\end{equation}

For every input dimension $p$, the model evaluates the Chebyshev polynomials from $T_{0}$ up to $T_{\mathrm{degree}}$.
Consequently, for each dimension $p$, we obtain a multi‑order feature set
$\bigl\{T_{0}(x_{p}),\,T_{1}(x_{p}),\,\dots,\,\allowbreak T_{K}( x_{p})\bigr\}, \;
$ where $K = degree.$

These polynomial features are subsequently combined with a set of learnable coefficients to produce the layer’s output.
In the original ChebyKAN implementation, all features generated by input dimension $p$ and polynomial order $K$ are concatenated to form a large feature vector:
\begin{equation}
h=[T_{0}(\tilde{x}_{1}),T_{1}(\tilde{x}_{1}),...,T_{K}(\tilde{x}_{1}),T_{0}(\tilde{x}_{2}),...,T_{K}(\tilde{x}_{n})].
\end{equation}

\begin{figure*}
  \centering
  \includegraphics[width=0.95\linewidth]{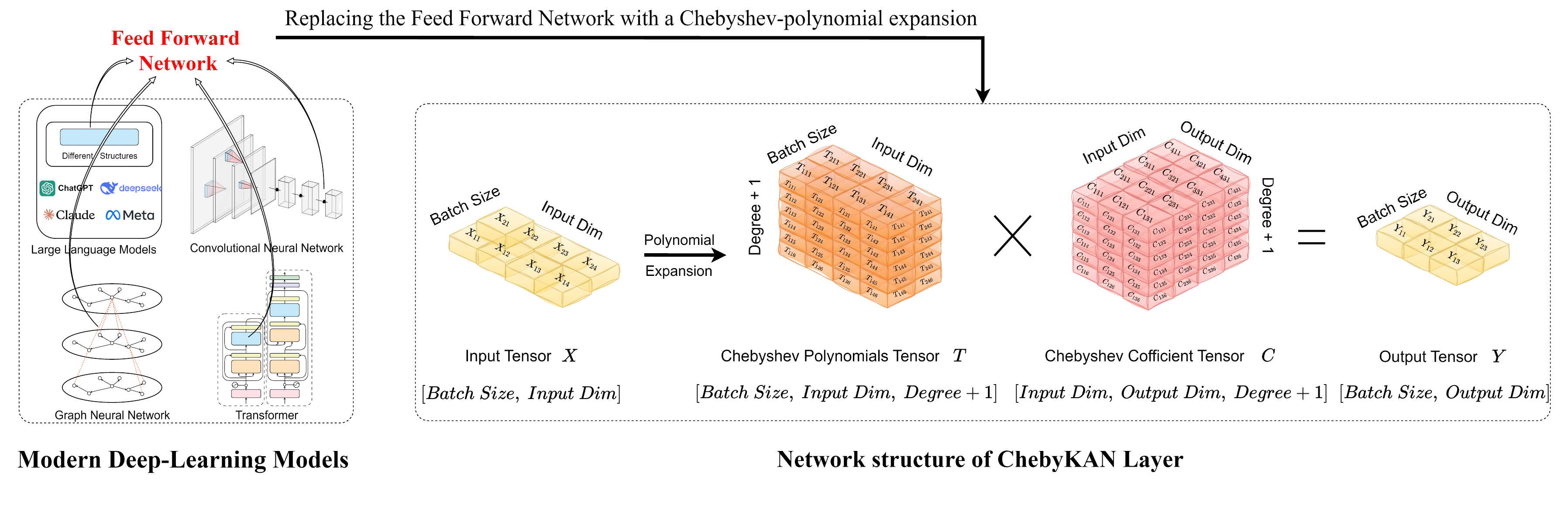}
  \caption{Replacing a conventional feed‑forward layer in deep‑learning models with a \textbf{ChebyKAN} layer: the input $X$ is mapped elementwise by the Chebyshev‑polynomial basis, producing the basis tensor $T$ with entries $T_{b,j,d}=T_d(tanh(X_{b,j}))$. The tensor $T$ is then linearly contracted with the learnable coefficient tensor $C$ to yield the output $Y$.}
  \label{fig_3}
\end{figure*}

Let $\mathbf{W}$ denote the learnable ``coefficient matrix'', the mapping can be written as: $y=W \cdot h +b,$
where $y$ denotes the output of the current layer. The network architecture of a ChebyKAN layer is shown in Figure \ref{fig_3}.

\subsection{The characteristics of the KAN variants}
\label{sec:kan_variants}
Numerous KAN variants share a common computational skeleton---generating multi-order basis functions for each input 
dimension and aggregating them with learnable coefficients.
Trigonometric-based forms (e.g., Chebyshev, Fourier) leverage recurrences to propagate orders without repeated 
\(\sin/\cos\) evaluations. For instance, FourierKAN can exploit identities such as $\cos\!\bigl((k+1)x\bigr)=\cos(kx)\cos(x)-\sin(kx)\allowbreak\sin(x),$ thereby propagating between successive orders without invoking $\sin(kx)$ or $\cos(kx)$ for every $k$. 

Orthogonal-polynomial and piecewise basis (e.g., Legendre, Hermite, B-splines) exhibit similar recurrence forms. Abstractly,
$
\alpha_k(x)\allowbreak \,B_{k+1}(x) \;=\; \beta_k(x)\,B_k(x)\;-\;\gamma_k\,B_{k-1}(x),
$
leads to similar dataflows and memory-access patterns during expansion and coefficient aggregation.

It is evident that despite differences in basis form and theoretical origin, most of the KAN variants share the consistent framework of multi‑order basis expansion and learnable coefficient aggregation. In the following, we take ChebyKAN as a representative case and conduct a detailed analysis of its performance bottlenecks.

\section{Performance Analysis and Motivation}

\label{cost}
In this section, we use \emph{ChebyKAN} as a representative to analyze why KAN variant operators underutilize GPUs and motivate our optimizations.
Adopting the Roofline perspective~\cite{williams2009roofline}, we balance compute and bandwidth to identify the dominant bottlenecks. The conclusions are also applicable to other KAN variants.

\begin{figure}[tp]
  \centering
  \includegraphics[width=0.88\linewidth]{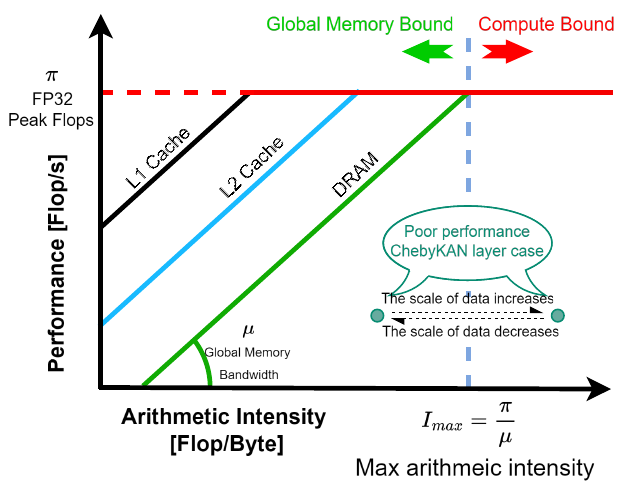}
  \caption{The roofline model of the Kolmogorov-Arnold network.}
  \label{fig_4}
\end{figure}

\subsection{Diagnosis of performance bottlenecks}
Although both the trigonometric Eq.~\eqref{th:chebyshev} and recurrence Eq.~\eqref{re:chebyshev} formulations of Chebyshev polynomials mentioned in \S \ref{fomula} are valid, the recurrence form is preferred for superior GPU efficiency. We analyze its performance bottlenecks on GPU hardware. The parameters and notations in this paper are listed in Table \ref{table_1}.

\begin{table}[htbp]
    \centering
    \caption{Configurations of ChebyKAN.}
    \begin{tabular}{ll}
        \toprule
        \textbf{Symbol} & \textbf{Meaning} \\
        \midrule
        $B$          & Batch size of input data \\[2pt]
        $D_{in}$     & Dimension of input data \\[2pt]
        $D_{out}$    & Dimension of output data \\[2pt]
        $d$     & Maximum order of a polynomial \\[2pt]
        $\lambda$    & Bytes per element \\[2pt]
        $TILE\_IN$    & Thread-block tile size along Input  \\[2pt]
        $TILE\_OUT$    & Thread-block tile size along Output  \\[2pt]
        $g\_x$   & Number of input tiles, $g_x=\bigl\lceil \tfrac{D_{\mathrm{in}}}{\texttt{TILE\_IN}}\bigr\rceil$\\[2pt]
        $g\_y$   & Number of output tiles, $g_y=\bigl\lceil \tfrac{D_{\mathrm{out}}}{\texttt{TILE\_OUT}} \bigr\rceil$\\[2pt]
        \bottomrule
    \end{tabular}
    \label{table_1}
\end{table}

As illustrated in Figure \ref{fig_3}, the forward propagation process of ChebyKAN can be roughly divided into two steps: 
calculating the polynomial values of all orders and multiplying-accumulating the polynomial expansion results with the learnable coefficient matrix.

Counting fused multiply-adds conservatively as \(2\) FLOPs, the layer’s total work and main data movement scale as
\begin{align*}
T & = T _{expand} +T _{combine}\;\approx\; 2B D_{\mathrm{in}}\!\left(d+(d{+}1)D_{\mathrm{out}}\right),\\
S &\;\approx\; \lambda\!\left[ B D_{\mathrm{in}} + B D_{\mathrm{out}} + 2B D_{\mathrm{in}}(d{+}1) + D_{\mathrm{in}}D_{\mathrm{out}}(d{+}1)\right].
\end{align*}
yielding arithmetic intensity:
\[
I \;=\; \frac{T}{S}
\;\approx\;
\frac{2B D_{\mathrm{in}}\!\left(d+(d{+}1)D_{\mathrm{out}}\right)}
{\lambda\!\left[B(D_{\mathrm{in}}{+}D_{\mathrm{out}})+2B D_{\mathrm{in}}(d{+}1)+D_{\mathrm{in}}D_{\mathrm{out}}(d{+}1)\right]}.
\]
where $I$ scales linearly with the batch size $B$ and the output dimension $D_{out}$. 
Consequently, as shown in Figure \ref{fig_4} \emph{ChebyKAN} operates in two distinct regimes:

\begin{itemize}
    \item \textbf{Memory‑bound ($I < I_{\text{max}}$).}  
      Arithmetic intensity becomes too low to saturate the GPU’s compute units. Each step of the recurrent formulation must access and update the result from the previous order. Unlike matrix multiplication, the required data cannot be pre‑loaded and processed in one highly parallel pass. Additionally, if the computation of small segments is not properly batched within the same block/warp, the kernel incurs frequent global‑memory accesses or redundant data loads.

    \item \textbf{Compute‑bound ($I \ge I_{\text{max}}$).}
      Scaling both $B$ and $D_{\text{out}}$ pushes the arithmetic intensity $I$ past the ridge point. 
      Yet the loop‑\allowbreak carried dependencies in the recurrent basis‑function generator restrict 
      instruction‑level parallelism. Additionally, the enlarged set of polynomials and coefficients inflate the per‑\allowbreak thread register footprint, thereby lowering Streaming Multiprocessor(SM) occupancy.
\end{itemize}

Therefore, our optimization strategy focuses on cutting global-memory traffic and maximizing in-block data reuse. These measures simultaneously raise arithmetic intensity in memory-bound case and improve compute utilization when the workload crosses into the compute-bound regime.



\subsection{Related GPU optimization work}
From the viewpoint of classical high‑performance computing (HPC) and numerical analysis, both industry and academia have already undertaken extensive optimization of \textbf{trigonometric function} and \textbf{polynomial recurrence}.  
Libraries such as Intel Vector Math Library (VML) \cite{greer2001scientific}, the CUDA math API \cite{cudaMathGuide}, and other specialized vector‑function packages offer hand‑tuned vectorization, SIMD\allowbreak /SIMT parallelism, and approximation techniques that greatly reduce the latency of single‑point or small‑batch function calls. For the aggregation phase, dense GEMM libraries such as cuBLAS and
CUTLASS attain near-peak hardware throughput. Some recent work, such as FusedFourierKAN \cite{FusedFourierKAN2024}, has optimized the performance of the KAN variant based on Fourier polynomial from an overall operator perspective.

However, these libraries perform well at their respective micro-kernels, with only particular execution stages optimized. 
Traditional vector math libraries usually only map functions like $\cos(x)$ and $\exp(x)$to vectorized implementations, 
and do not fully optimize for the large number of intermediate steps that exist in multi-order recurrence. 
They also lack a design that fuses these basis function computations with the subsequent multiply-accumulate of learnable coefficients. Although FusedFourierKAN attempts to fill this gap, it relies on simple kernel fusion, yielding limited performance gains, and its design is tightly coupled to the Fourier basis, which hinders extension to other polynomial bases. This motivates the general optimization pipeline proposed in the present work. To our knowledge, our proposed solution first provides a unified, variant-agnostic acceleration framework for KAN-style operators.


\section{Methodology}

\subsection{Overview}
To accelerate both forward and backward propagation of KAN operators
(e.g., \textsc{ChebyKAN}, \textsc{LegendreKAN}) on modern GPUs, we design a generic optimization pipeline. Its guiding principle is to maximize parallelism while minimizing memory pressure without sacrificing support for different polynomial bases. The overall design is shown in Figure \ref{architecture}. The proposed pipeline consists of four orthogonal methods:

\begin{figure}[htbp]
  \centering
  \includegraphics[width=0.95\linewidth]{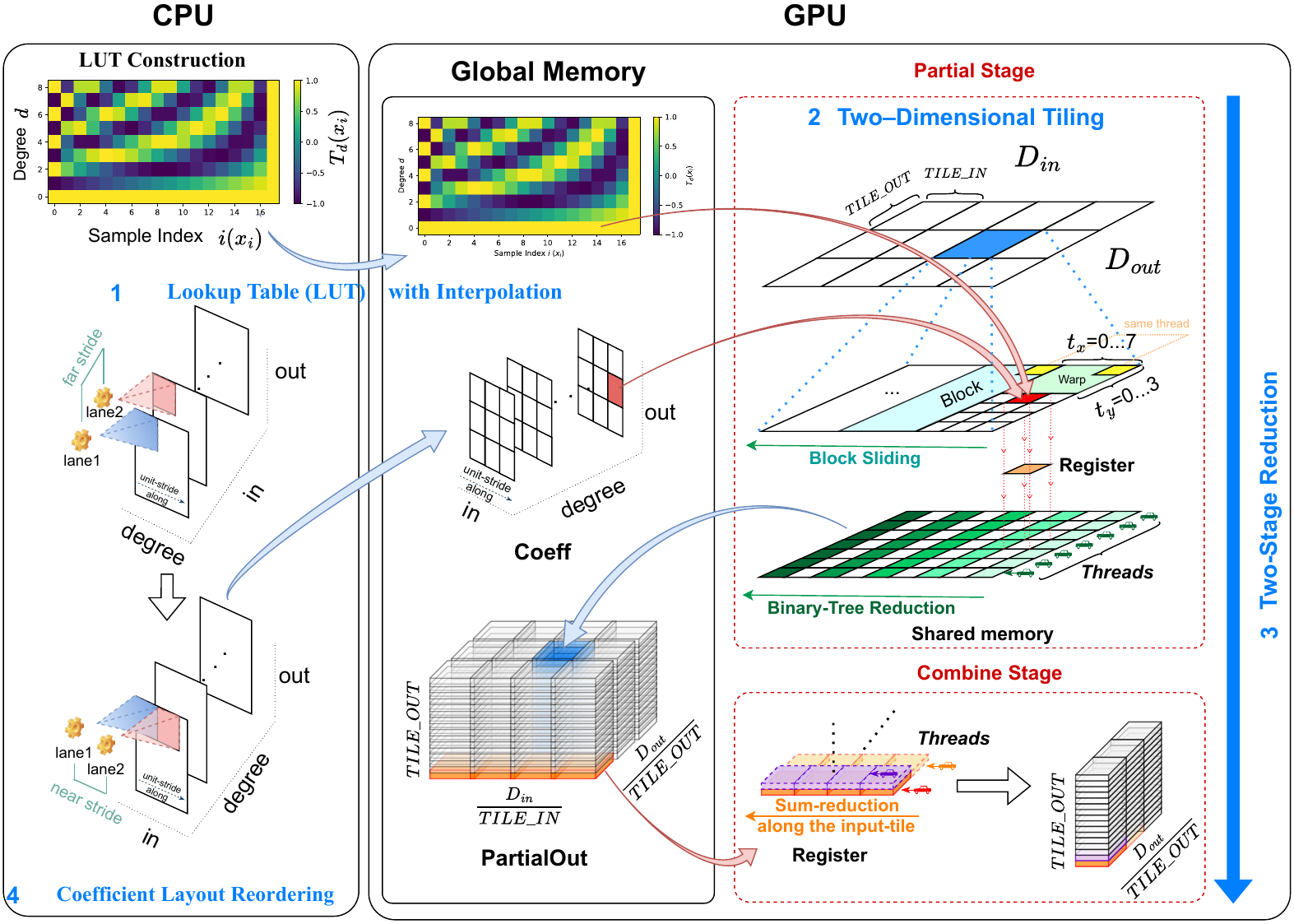}
  \caption{The overall design of the KAN variant acceleration.}
  \label{architecture}
\end{figure}

\begin{enumerate}
  \item \textbf{Lookup Table (LUT) with Interpolation}.  
        The basis functions of many polynomials (e.g.,\ Chebyshev, Legendre)
        can be pre-computed offline and stored in a large LUT \cite{aus2006table}. At run time, we obtain approximations by linear (or higher-order) interpolation,  eliminating expensive trigonometric evaluations or recurrence formulations. Our implementation allocates the LUT in global memory, so as to meet high-precision requirements while alleviating constant-memory capacity constraints.

  \item \textbf{2D Tiling}.  
        We adopt a 2D tiling strategy by simultaneously partitioning the input and output dimensions into rectangular blocks of configurable size. Each GPU thread block is assigned to process a single tile, performing the corresponding multiply-accumulate operations locally. This design improves data access spatial locality, which enhances cache reuse and enables fine-grained parallelism across both dimensions.

  \item \textbf{Two-Stage Reduction (\emph{Partial} + \emph{Combine})}.  
        We adopt a two-stage scheme to avoid large-scale \texttt{atomicAdd} operations on the same output location which cause severe resource contentions. In the \emph{Partial} stage, each tile accumulates its partial sum in shared memory. The \emph{Combine} stage then merges partial results from different tiles into the final output, reducing atomic contention and write-conflict overhead.

  \item \textbf{Coefficient Layout Reordering}.  
        The original coefficient tensor is usually stored as
        $[\mathit{inputdim},\,\mathit{outputdim},\,\allowbreak \mathit{degree}+1]$, which leads to large access strides inside the kernel. We reorder it to
        $[\mathit{degree}+1,\,\mathit{outputdim},\,\allowbreak \mathit{inputdim}]$, enabling
        contiguous memory accesses and higher bandwidth utilization.
\end{enumerate}

By applying the four key optimization steps described above, the proposed implementation significantly reduces the number of explicit function calls, bandwidth waste, and atomic collisions in both forward and backward propagation.

\subsection{Polynomial Operators Acceleration via LUT}
\label{LUT}
Many polynomial basis functions employed in the KAN variants, such as Chebyshev, Legendre, and Hermite, share two universal properties:

\begin{itemize}
  \item \textbf{Domain normalization to \([-1,1]\).}  
        Each basis is either intrinsically defined on the interval \([-1,1]\) or can be mapped to that interval by a simple normalization step. For example, the input \(x\) in ChebyKAN can be transformed by \(\tanh(\cdot)\) so as to ensure that \(x\in[-1,1]\).

  \item \textbf{Offline discretization and storage.}  
        The polynomial function \(p_d(x)\) of any KAN variant attains a deterministic value for fixed \(degree\) and \(x\). Therefore, we can sample \(p_d(x)\) on the CPU over the interval \([-1,1]\) with an appropriate step size, compute the results iteratively, and store them in a LUT.
\end{itemize}

These properties enable the \emph{LUT-interpolation} strategy and demonstrate its \emph{generality}:
regardless of the particular polynomial basis, once the function can be discretized over \([-1,1]\) and a LUT built for the required \(\mathit{degree}\), the same strategy is always applicable.

\subsubsection{Offline construction of LUT}\leavevmode\\
We first construct the LUT on the CPU. For a prescribed maximum \(\mathit{degree}\) 
we choose a table size \(\mathit{LUT\_SIZE}\) and discretize the interval \([-1,1]\) with the uniform step:
$
  \Delta \;=\; \frac{2}{\mathit{LUT\_SIZE}-1}.
$

At each grid point \(x_i = -1 + i\,\Delta\) (\(i = 0,1,\ldots,
\mathit{LUT\_SIZE}-1\)) we evaluate the sequence
\(T_{0}(x_i), T_{1}(x_i), \ldots, T_{\mathit{degree}}(x_i)\) by applying the
recurrence in Eq.~\eqref{re:chebyshev} once, and write the results into a
two-dimensional array \(\mathbf{LUT}\).
\(LUT[d,i]\) stores the value of the \(d\)-th basis function at the
\(i\)-th sample. For other KAN variants, one merely replaces the Chebyshev recurrence with the corresponding polynomial relation, leaving the subsequent interpolation logic unchanged.

After LUT has been generated on the CPU, it is uploaded to the GPU.
While storing the table in read-only on-chip memory could reduce access latency, its size exceeds the available capacity. Consequently, the LUT is stored in global memory, providing every thread in the subsequent kernels with read-only access to the precomputed polynomial values.

\subsubsection{Online interpolation from LUT}\leavevmode\\
Once a GPU thread receives an input value \(x\in[-1,1]\), it approximates \(T_{d}(x)\) 
from the lookup table in two steps: (i) \emph{calculate its normalized position in the interval \([-1,1]\) based on \(x\):}
$pos \;=\; \frac{x+1}{2}\bigl(\mathit{LUT\_SIZE}-1\bigr);$ (ii)
\emph{perform linear interpolation} between the two neighbouring
        samples whose indices are \(\lfloor\texttt{pos}\rfloor\) and
        \(\lfloor\texttt{pos}\rfloor+1\).


This procedure produces a close approximation to \(T_{d}(x)\) without invoking run-time trigonometric functions or the recurrence relation.

Linear interpolation suffices for most applications: the grid spacing \(\Delta = 2/(\mathit{LUT\_SIZE}-1)\) is small with a large \(\mathit{LUT\_SIZE}\), rendering the interpolation error negligible over the interval. Higher-order schemes (quadratic or
cubic) could further reduce error but at the cost of additional arithmetic. Therefore, the linear variant strikes a favorable balance between simplicity and efficiency.

In both forward and backward propagation, both polynomial values and their derivatives can be obtained directly from the LUT. For the derivative, we can approximate \(\frac{d}{dx}T_{d}(x)\) by a finite difference between neighbouring table entries, or store the pre-computed differences as an auxiliary LUT. In this work, using the Chebyshev basis as an example, the backward gradient is computed from either \(T_{\text{approx}}\) or the difference of adjacent samples, avoiding explicit evaluations of the analytic derivative. \S \ref{tile} revisits this strategy when discussing backward propagation \ref{alg:backward-tiling}.

\subsection{Parallelism of 2D tiling across the input and output dimensions} \label{tile}

The computational workload scales with \(D_{\mathrm{in}}\cdot D_{\mathrm{out}}\). One-dimensional parallelization over the batch axis forces warps to stride long distances in either dimension, leading to non-coalesced memory traffic. We therefore adopt \textbf{2D tiling}: partition input into \textit{TILE\_IN} and output into \textit{TILE\_OUT}. Each CUDA block is assigned a \(\textit{TILE\_IN}\times\textit{TILE\_OUT}\) sub-matrix, but it processes it using a specialized ``output-aligned'' warp strategy. This design (i) improves spatial locality for both coefficient and LUT accesses, (ii) confines \texttt{atomicAdd} operations to localized output regions, reducing contention, and (iii) generates a larger set of load-balanced blocks, enabling the scheduler to saturate all SMs more effectively.

\textbf{Block--grid configuration and thread mapping.} 
Let \(\text{grid}=\Bigl(g_x,\ g_y,\ B\Bigr)\), where \(g_x=\bigl\lceil \tfrac{D_{\mathrm{in}}}{\texttt{TILE\_IN}} \bigr\rceil\) and \(g_y=\bigl\lceil \tfrac{D_{\mathrm{out}}}{\texttt{TILE\_OUT}} \bigr\rceil\). Critically, we set the block dimensions to \texttt{block} = (\textit{BLOCK\_ \allowbreak DIM\_X}, \textit{BLOCK\_DIM\_Y}), where \(\textit{BLOCK\_DIM\_Y}\) is typically 32 (a full warp) and \(\textit{BLOCK\_DIM\_X}\) is a smaller value (e.g., 8). A block \((\mathrm{tileI},\mathrm{tileO},b)\) processes the sub-tile \((\Delta_i, \Delta_o)\) using a different thread mapping:
\begin{itemize} 
    \item The thread's \texttt{y} index, \(t_y\), maps directly to the output dimension: \(\mathrm{o}=\Delta_o\!.\text{start}+t_y\). This aligns a full warp (32 threads) along the output dimension.
    \item The thread's \texttt{x} index, \(t_x\), maps to the \textit{offset} of the input dimension \textit{j}. 
\end{itemize}

\begin{figure}[tb]
  \centering
  \includegraphics[width=0.9\columnwidth]{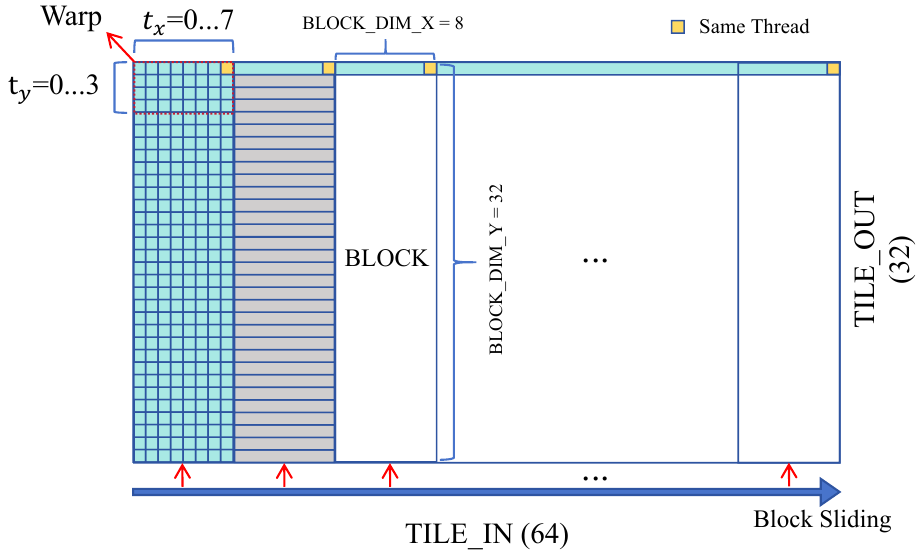}
  \caption{
    Visualization of the output-aligned 2D tiling strategy. A single BLOCK (8$\times$32 threads) maps its ty index 1:1 to the TILE\_OUT dimension.
    A hardware Warp is an 8$\times$4 tile. The "Same Thread" (yellow) iterates 
    across the TILE\_IN dimension with a stride of BLOCK\_DIM\_X (8).
  }
  \label{fig:tiling_diagram} 
\end{figure}

To cover the entire \textit{TILE\_IN} range (e.g., 64), each thread is assigned multiple \textit{j} indices to process. This iteration uses the thread's \(t_x\) index as an initial offset and a stride of \textit{BLOCK\_\allowbreak DIM\_X}. Figure \ref{fig:tiling_diagram} provides a visual representation of this block configuration and iterative mapping.

This \textit{output-aligned} configuration is a deliberate choice to resolve a critical performance bottleneck in the \texttt{LUT} access. This bottleneck arises from a fundamental trade-off between the ideal access patterns for the \texttt{LUT} and \texttt{Coeff} tensors:
\begin{itemize}
    \item \textbf{Ideal \texttt{LUT} Access (1-way Broadcast):} To minimize \texttt{LUT} memory divergence, a warp requires as few distinct \textit{j} values as possible. The ideal scenario provides \textbf{1 distinct \textit{j} value} and allows for a perfect 1-way broadcast from the \texttt{LUT}.
    \item \textbf{Ideal \texttt{Coeff} Access (32-way Coalescing):} To maximize \texttt{Coeff} coalescing, a warp requires many consecutive \textit{j} values. The ideal scenario provides \textbf{32 distinct \textit{j} values} for a perfect 32-way coalesced access.
\end{itemize}

These two ideal scenarios are mutually exclusive.
A naive tiling strategy (e.g., a \(64 \times 16\) block mapping \texttt{tx->j}) would result in a hardware warp containing 32 distinct \textit{j} values. Since the \texttt{LUT} index \texttt{idx} is calculated from \textit{j}, this naive mapping causes a severe \textbf{32-way memory scatter}.

Our \((8, 32)\) block design is specifically chosen to be a trade-off. A warp in this configuration is an \(8 \times 4\) tile, containing only \textbf{8 distinct \textit{j} values} (one for each \(t_x\)). This design reduces the LUT memory bottleneck to a much more manageable \textbf{8-way scatter} when accessing the \texttt{LUT}.

\textbf{Coefficient/LUT memory layout and coalesced accesses}.
As established in the previous section, our \((8, 32)\) block configuration is designed to solve the primary bottleneck of \texttt{LUT} memory by reducing it to an 8-way scatter. This block-level decision, in turn, dictates the optimal memory layout for the \texttt{Coeff} tensor.

Given this \((8, 32)\) block , a hardware warp is an \(8 \times 4\) tile, and its 8 \textit{consecutive} threads (e.g., \mbox{\texttt{ty=0, tx=0..7}}) are mapped to 8 different \textit{j} indices . We must choose a layout that makes these 8 accesses efficient.

Therefore, we reorder the coefficient tensor to the \mbox{\([d, o, j]\)} layout , where \textit{j} is the \textbf{innermost dimension}. With this layout, the same 8 consecutive threads now access contiguous memory addresses. Therefore, the hardware executes this as a single, efficient \textbf{8-way coalesced read}. This same coalescing benefit applies to the \texttt{atomicAdd} operations during the backward pass. This strategy, detailed further in \S \ref{rearrangement}.

\textbf{The forward and backward propagation process.} In the forward propagation, each sub-block carries out the multiply-\allowbreak accumulate of polynomial values over its designated $\Delta_i$ and $\Delta_o$. The forward propagation can be summarized by Algorithm \ref{alg:forward-tiling}, which reflects the new outer iterative loop over the \textit{j} dimension. Each thread evaluates Chebyshev terms via \texttt{LookupCheby} on $\tanh(x_{b,j})$ (§\ref{LUT}) and accumulates over the degree in registers, returning the vector \(T_{0}(x_i), T_{1}(x_i),\dots, \allowbreak T_{\mathit{degree}}(x_i)\).

During the backward propagation, we apply the same output-aligned 2D tiling strategy to the output gradient \(\partial\mathcal{L}/\partial\mathbf{y}\). Within each \((\Delta_i,\Delta_o)\) sub-tile, the gradients with respect to both the coefficients and the inputs are computed in batch. The corresponding pseudocode is presented as Algorithm \ref{alg:backward-tiling}.

\begin{algorithm}[tb]
\normalsize
\DontPrintSemicolon
\SetAlgoVlined

\SetKwFunction{Lookup}{LookupCheby}
\SetKwFunction{min}{min}
\SetKwFunction{Z}{Zero}
\SetKwFunction{tanh}{tanh}
\SetKwFunction{dot}{dot}

\DontPrintSemicolon
\KwIn{\,\,\,\,\,$X \!\in\! \mathbb{R}^{B \times D_{\text{in}}}$,\,
      $Coeff \!\in\! \mathbb{R}^{(\mathit{degree}+1) \times D_{\text{out}} \times D_{\text{in}}}$,\; \,\,\,\,\,\,\,\,\,\,\,\,\,\,\,\,\,\,\,\,\,\,
      $LUT \!\in\! \mathbb{R}^{(\mathit{degree}+1) \times \textit{LUT\_SIZE}}$, \;
      \,\,\,\,\,\,\,\,\,\,\,\,\,\,\,\,\,\,\,\,\,\,
      $tile_{\mathrm{in}}$,\, $tile_{\mathrm{out}}$,\, $BLOCK\_DIM\_X$}
\KwOut{\,$\mathit{partialOut} \!\in\! \mathbb{R}^{(g_x \times g_y \times B \times \textit{tile\_out})}$}

\For{$b \leftarrow 0$ \KwTo $ B-1 $ \KwSty{in parallel}}{
    \For{$tileI \leftarrow 0$ \KwTo $\lceil D_{\text{in}}/\textit{tile\_in}\rceil - 1$ \KwSty{in parallel}}{
        \For{$tileO \leftarrow 0$ \KwTo $ \lceil D_{\text{out}}/\textit{tile\_out}\rceil - 1$ \KwSty{in parallel}}{
            $in_{start} \leftarrow tileI \times \textit{tile\_in}$\;
            $in_{end}   \leftarrow \min(in_{start} + \textit{tile\_in},\, D_{\text{in}})$\;
            $out_{start}\leftarrow tileO \times \textit{tile\_out}$\;
            $out_{end}  \leftarrow \min(out_{start} + \textit{tile\_out},\, D_{\text{out}})$\;
            $g_x \leftarrow \lceil D_{\text{in}}/\textit{tile\_in}\rceil$\;

            $\mathit{smem}[BLOCK\_DIM\_Y][BLOCK\_DIM\_X]$\;

            \For{$(t_x, t_y) \leftarrow (0..BLOCK\_DIM\_X-1, 0..BLOCK\_DIM\_Y-1)$ \KwSty{in parallel}}{
                $\mathit{out} \leftarrow out_{start} + t_y$\;
                $\mathit{sum\_total} \leftarrow 0$\;

                \If{$\mathit{out} < out_{end}$ \KwSty{AND} $b < B$}{
                    \For{$j \leftarrow in_{start} + t_x$ \KwTo $in_{end} - 1$ \KwSty{step} $BLOCK\_DIM\_X$}{
                        $x \leftarrow \tanh{X[b,j]}$\;
                        $\mathit{T_{\text{approx}}} \leftarrow \Lookup{LUT,\, x,\, \textit{degree}}$\;
                        $\mathit{sum\_total} \leftarrow \mathit{sum\_total} + dot(Coeff[:,\,\mathit{out},\,j], \, \mathit{T_{\text{approx}}})$\;
                    }
                }
                $\mathit{smem}[t_y, t_x] \leftarrow \mathit{sum\_total}$\;
                \text{WaitAndReduceOverX}($\mathit{smem}[t_y], t_x$)\; 
                \If{$t_x = 0$ \KwSty{AND} $\mathit{out} < out_{end}$ \KwSty{AND} $b < B$}{
                   $\mathit{gIdx} \leftarrow (tileO \times g_x + tileI) \times B \times \textit{tile\_out} + b \times \textit{tile\_out} + t_y$\;
                   $\mathit{partialOut}[\mathit{gIdx}] \leftarrow \mathit{smem}[t_y, 0]$\;
                }
            }
        }
    }
}
\caption{Forward propagation (Partial Stage) with output-aligned 2D tiling and \mbox{\([d, o, j]\)} layout.}
\label{alg:forward-tiling}
\end{algorithm}

\begin{algorithm}[tb]
\normalsize
\DontPrintSemicolon
\SetAlgoVlined
\SetKwComment{Comment}{$\triangleright$ }{}

\SetKwFunction{Lookup}{LookupChebyAndDiff}
\SetKwFunction{min}{min}
\SetKwFunction{tanh}{tanh}

\KwIn{\,\,\,\,\,$X \!\in\! \mathbb{R}^{B \times D_{\text{in}}}$,\,
      $Coeff \!\in\! \mathbb{R}^{(\mathit{degree}+1) \times D_{\text{out}} \times D_{\text{in}}}$,\; \,\,\,\,\,\,\,\,\,\,\,\,\,\,\,\,\,\,\,\,\,\,
      $LUT \!\in\! \mathbb{R}^{(\mathit{degree}+1) \times \mathit{LUT\_SIZE}}$, \;
      \,\,\,\,\,\,\,\,\,\,\,\,\,\,\,\,\,\,\,\,\,\,
      $\mathit{tile_{\mathrm{in}}}$,\, $\mathit{tile_{\mathrm{out}}}$,\, $\mathit{BLOCK\_DIM\_X}$}
\KwOut{\,$\mathit{coeff\_grad} \!\in\! \mathbb{R}^{(\mathit{degree}+1)\times D_{\text{out}}\times D_{\text{in}}}$,\;
      \,\,\,\,\,\,\,\,\,\,\,\,\,\,\,\,\,\,\,\,\,\,
       $\mathit{x\_grad} \!\in\! \mathbb{R}^{B \times D_{\text{in}}}$}

\tcp{The following is executed by each block $(tileI, tileO, b)$ in parallel}
$\mathit{in_{start}} \leftarrow \mathit{tileI} \times \mathit{tile\_in}$\;
$\mathit{in_{end}}   \leftarrow \min(\mathit{in_{start}} + \mathit{tile\_in},\, D_{\text{in}})$\;
$\mathit{out_{start}}\leftarrow \mathit{tileO} \times \mathit{tile\_out}$\;
$\mathit{out_{end}}  \leftarrow \min(\mathit{out_{start}} + \mathit{tile\_out},\, D_{\text{out}})$\;
$\mathit{smemX}[\mathit{BLOCK\_DIM\_X}][\mathit{BLOCK\_DIM\_Y}]$\; \tcp{Shared mem for x\_grad}

\tcp{Each thread $(t_x, t_y)$ executes in parallel}
\If{$b < B$}{
    \For{$j \leftarrow \mathit{in_{start}} + t_x$ \KwTo $\mathit{in_{end}} - 1$ \KwSty{step} $\mathit{BLOCK\_DIM\_X}$}{
        $\mathit{out} \leftarrow \mathit{out_{start}} + t_y$\;
        $\mathit{x\_grad\_partial} \leftarrow 0$\;
        
        \If{$\mathit{out} < \mathit{out_{end}}$}{
            $g_o \leftarrow dY[b, \mathit{out}]$\;
            $x \leftarrow \tanh(X[b,j])$\;
            $(\mathit{T_{\text{approx}}},\,\mathit{dTdx}) \leftarrow \Lookup(LUT,\,x,\,\mathit{degree})$\;
            $\mathit{sum\_dx} \leftarrow 0$\;
            
            \For{$d \leftarrow 0$ \KwTo $\mathit{degree}$}{
                $\text{atomicAdd}(\mathit{coeff\_grad}[d,\,\mathit{out},\,j],\, g_o \times \mathit{T_{\text{approx}}}[d])$\;
            }
            
            \For{$d \leftarrow 1$ \KwTo $\mathit{degree}$}{
                $\mathit{c\_val} \leftarrow Coeff[d,\,\mathit{out},\,j]$\;
                $\mathit{sum\_dx} \leftarrow \mathit{sum\_dx} + g_o \times \mathit{c\_val} \times \mathit{dTdx}[d]$\;
            }
            $\mathit{x\_grad\_partial} \leftarrow \mathit{sum\_dx}$\;
        } 
        
        $\mathit{smemX}[t_x, t_y] \leftarrow \mathit{x\_grad\_partial}$\;
        $\mathit{final\_x\_sum} \leftarrow \text{BlockReduceOverY}(\mathit{smemX}[t_x, :])$\; \tcp{Syncs \& sums smemX[tx][0..31]}
        
        \If{$t_y = 0$}{
            $\text{atomicAdd}(\mathit{x\_grad}[b,\,j],\, \mathit{final\_x\_sum})$\;
        }
        $\text{SyncBlock}()$\; \tcp{Wait for all threads before next j}
    } 
} 

\caption{Backward propagation with output-aligned 2D tiling and \mbox{\([d, o, j]\)} layout.}
\label{alg:backward-tiling}
\end{algorithm}


\subsection{Two-Stage Reduction}
\label{2stage}
When either the data dimensions or the batch size are large, each CUDA block must accumulate a portion of the output (or its gradient) within a single kernel launch. Even with 2D tiling along the input and output axes, writing directly to the global output or the gradient array still incurs three major drawbacks:

\begin{itemize}
  \item \textbf{Atomic contention.}  
        Whenever multiple blocks update the same batch item or the same output element, a large number of \texttt{atomicAdd} operations are inevitable and quickly become a performance bottleneck.

  \item \textbf{Write-after-read latency.}  
        Since atomic updates are serialized by the hardware, repeated atomic writes to the same address within one kernel stall the instruction pipeline and introduce extra latency.
  
  \item \textbf{Superfluous atomic overhead.}  
        Atomic writes are significantly more expensive than ordinary stores. Aggregating partial sums in an intermediate buffer and performing a single consolidated write-back would cut down the total number of atomic operations.
        
\end{itemize}

To mitigate these issues we introduce a \textbf{two-stage reduction} (\emph{Partial} + \emph{Combine}). In the \emph{Partial} stage, each block stores its results in a private, conflict-free buffer. Subsequently, in the \emph{Combine} stage, these buffers are merged into the global output, converting many fine-grained atomic updates into a small, well-controlled batch of writes.

\subsubsection{Partial and Combine stages}\leavevmode\\
In forward propagation, each thread block writes its partial sum to device-global workspace buffer \texttt{PartialOut} 
instead of issuing global atomics. Every sub-block created by the 2D tiling scheme owns a dedicated segment 
(denoted by \texttt{subBlockID}) inside this buffer, preventing write conflicts between blocks.

After completion of the \emph{Partial} stage, the partial sums generated by all sub-blocks reside in \texttt{PartialOut}. 
The subsequent \textbf{Combine} stage merges these values. This process is typically completed within a single kernel launch, 
obtaining the final output \(\mathbf{y}\).





Since the values can be accumulated sequentially while iterating over the \texttt{subBlockID}s, the amount of write contention is greatly reduced at this stage. If a fully parallel merge is desired, hierarchical or tree-based reduction schemes can be employed to balance efficiency against implementation complexity.

\subsubsection{Performance benefit analysis}\leavevmode\\
Compared to a single-kernel direct write-back, the two-stage reduction maintains a unique-writer invariant for every intermediate and final location, thus eliminating all global atomics in forward. The Combine stage performs ordinary streaming loads and a single store per \((b, out)\). For gradients, we retain a minimal set of atomics: (i)coefficient gradient aggregates across batches on the same \((d, j, out)\); (ii)input gradient uses per-block reduction along \(out\), followed by a single atomic add per \((b, j)\) to merge contributions across \(g_y\) output tiles. This reduces atomics from \(BD_{in}D_{out}\) to \(BD_{in}g_y\).

\paragraph{Atomic update count (forward).}
A block-reduction + atomic-write baseline performs
$N^{\mathrm{base}}_{\mathrm{atomic,fwd}}=B\cdot D_{out}\cdot g_x$ atomic updates.
Our two-stage reduction eliminates them:
$N^{\mathrm{ours}}_{\mathrm{atomic,fwd}}=0$.

\paragraph{Bandwidth and temporary footprint.}
Two-stage introduces a streaming intermediate
$S_{\mathrm{partial}}\approx \lambda\,B\,D_{out}\,g_x$ bytes and extra I/O
$2 \times S_{\mathrm{partial}}(\text{write}+\text{read})$.
Two-stage is beneficial when $g_x\,c_a \gtrsim g_x(c_r+c_w)+c_w$,
i.e., when the per-atomic cost $c_a$ dominates normal read cost $c_r$ and write costs $c_w$.

\paragraph{Backward $x$-gradient.}
A naive design would require $B\cdot D_{in}\cdot D_{out}$ atomics.
With intra-block reduction along \textit{out}, ours reduces it to
$N^{\mathrm{ours}}_{\mathrm{atomic,bwd-x}}=B\cdot D_{in}\cdot g_y$,
avoiding fine-grained atomic contention. A further two-stage process would remove
atomics at the cost of an extra buffer of $\lambda \,B\,D_{in}\,g_y$ bytes and an
additional launch, which is not cost-effective in our methodology.

\subsection{Coefficient layout reordering}
\label{rearrangement}
In typical implementations of KAN networks, the polynomial coefficients are stored for every input dimension~\(\mathit{j}\) and output dimension~\(\mathit{o}\). When the tensor is kept in its default three-dimensional layout, GPU threads must access elements with excessively large strides, which in turn diminishes both coalesced memory throughput and cache efficiency.

\subsubsection{Original layout and memory-access pattern}\leavevmode\\
In the baseline implementation, the coefficients are stored in the index order \((j,o,d)\), i.e.\ as a three-dimensional array \texttt{coeff[$D_{\text{in}}$]\allowbreak [\,$D_{\text{out}}$][\,$\mathit{degree}{+}1$]}. Although this layout is intuitive for CPU code --- one loops over every input index~\(j\) and output index~\(out\) before selecting the desired degree~\(d\), it leads to sub-optimal access patterns on the GPU. Under a 2D tiling scheme in which a thread block sweeps across \(D_{\text{in}}\!\times\!D_{\text{out}}\) and then iterates over~\(d\), two contrasting stride behaviours emerge:

\begin{itemize}
\item When a \emph{single} thread reads successive polynomial orders \(\mathbf{coeff}[j,o,d]\), the stride along the \(d\)-dimension is small and cache-friendly.
\item However, if threads in the same warp handle neighbouring \((j,o)\) pairs, their memory accesses are scattered across the first two dimensions of the array, producing non-contiguous addresses and poor coalescing efficiency.
\end{itemize}

\subsubsection{Reordering the coefficient to the layout $[d,o,j]$}\leavevmode\\
To solve the access pattern problem, we reorder the coefficient tensor as $\texttt{coeff\_rearr}[d,o,j]$. This layout is not chosen in isolation; it is specifically designed to work in synergy with the \((8, 32)\) \textit{output-aligned} block strategy.

As established in \S \ref{tile}, our \((8, 32)\) block is designed to have 8 distinct \textit{j} values within a warp (from \(t_x=0..7\)) to solve the \texttt{LUT} bottleneck. With this block shape, a hardware warp is an \(8 \times 4\) tile, and its 8 \textit{consecutive} hardware threads (e.g., \mbox{\texttt{ty=0, tx=0..7}}) are mapped to 8 different but consecutive \textit{j} indices. We must choose a layout that makes these 8 simultaneous accesses efficient. Therefore, we reorder the coefficient tensor to the \mbox{\([d, o, j]\)} layout, where \textit{j} is the \textbf{innermost dimension}. Since these threads access contiguous memory addresses, the hardware executes this as a single, efficient \textbf{8-way coalesced read}.

Additionally, this layout benefits for both forward and backward propagations. Forward propagation reads \(\texttt{coeff}[d,\allowbreak o,j]\) with coalesced loads. Backward propagation writes to \(\partial\mathcal{L}\allowbreak /\partial \texttt{coeff}[d,o,j]\) , which benefits equally from \textbf{coalesced \texttt{atomicAdd}} operations, significantly reducing atomic contention on the memory bus.

When \(D_{\mathrm{in}}\) and \(D_{\mathrm{out}}\) are large, this new layout, combined with our specific tiling strategy, raises effective bandwidth and hence throughput on identical hardware.

Since this method relies only on the enumerability of the \(\mathit{degree}+1\) axis, it is applicable to Legendre, Hermite, and other KAN variants.

\begin{table*}[]
\centering
\caption{Benchmark Suite Summary}
\label{tab:benchmark_summary}
\renewcommand{\arraystretch}{1} 
\begin{tabular}{@{}l l l c c c@{}}
\toprule
\textbf{Dataset} & \textbf{Description} & \textbf{Layer Widths} & \textbf{Degree} & \textbf{Batch Size} \\
\midrule
Google Speech Commands v2& 105,872 utterances, 35 labels& $40 \rightarrow 256 \rightarrow 256 \rightarrow 12$ & 8 & 128 \\

\addlinespace 
VoiceBank-DEMAND & 28 speakers, 13 noises, 13 categories& $257 \rightarrow 512 \rightarrow 512 \rightarrow 13$ & 15 & 64 \\

\addlinespace
Kaggle House-Prices & 1460 properties, 79 raw features & $512 \rightarrow 1024 \rightarrow 1024 \rightarrow 1$ & 24 & 32 \\

\bottomrule
\end{tabular}
\end{table*}

\section{Experiment}
This section evaluates the practical effectiveness of the \allowbreak ChebyKAN optimizations. 
The study adopts a macro-to-micro protocol. Firstly, for three representative end-to-end workloads, we measure epoch-level training and inference latency as well as sample throughput 
for seven kernel versions, including both baselines, on an NVIDIA A100 GPU. Second, we isolate a single ChebyKAN layer under the same three input-output-\allowbreak degree configurations and record forward and backward latency to analyse operator-level scalability. Finally, we construct a roofline model to quantify how the proposed methodology schemes mitigate micro-architectural bottlenecks. 

Importantly, our goal is to evaluate the effectiveness of our proposed KAN operator as an MLP replacement. Thus, we choose ChebyKAN-dominant models to highlight the operator's performance gains. This approach is crucial for isolating the acceleration of our fused kernels, ensuring that the measured speedups are not confounded by other complex operators (e.g., attention mechanisms, graph convolutions or PDE residual calculations) present in larger, state-of-the-art architectures. Our ultimate aim is to validate that our optimized operator can serve as an efficient, plug-in replacement for \texttt{Linear} layers, enabling future work to reconstruct existing deep learning models with KAN variants.

\subsection{Hardware \& Software Environment}
All experiments are conducted on a single NVIDIA A100-SXM4-40 GB GPU which sustains theoretical peaks of 19.5 TFLOPS FP32.

The software stack comprises Ubuntu 22.04 LTS (kernel 5.15), CUDA 12.6. High-level execution 
relies on PyTorch 2.2.0+cu126. Micro-architectural data are gathered with Nsight Compute 2024.1 in kernel 
performance-replay mode. Timing is obtained with 
CUDA events placed around each kernel, synchronising the stream before and after measurement; 
data-loader overheads are excluded.  

\subsection{Workloads and Baselines}
\subsubsection{Benchmark Suite} \leavevmode\\
Our benchmark suite comprises three distinct datasets, with key parameters summarized in Table \ref{tab:benchmark_summary}. 
For \textbf{Google Speech Commands v2} \cite{warden2018speech}, each sample is a one-second utterance intended for single-word classification. 
The \textbf{VoiceBank-\allowbreak DEMAND} \cite{valentini2017noisy} dataset is constructed by mixing clean utterances from the VoiceBank corpus with various environmental noise recordings from the DEMAND database. It is repurposed for an acoustic-scene classification task. 
For the \textbf{Kaggle House-Prices} \cite{kaggle_houseprices} regression task, 
the model predicts the target market price from 79 raw features. The input is either zero-padded or truncated 
to a uniform dimension. 

These three tasks jointly cover audio classification, speech enhancement, and structured-data regression, 
and span operator \allowbreak widths from $40$ to $1024$, providing a representative test bed for 
assessing ChebyKAN performance.




\subsubsection{Baseline Implementations} \leavevmode\\
Seven kernels are evaluated. All kernels are functionally identical (FP32 forward, FP32 gradients). 
The description of each method is given in Table \ref{table_method}.

\begin{table*}[htbp]
    \centering
    \caption{Kernel version of baselines and our KAN implementations.}
    \begin{tabularx}{\textwidth}{|C{2.5cm}|X|X|}
        \hline
        \textbf{Kernel version} & \textbf{Incremental Optimization Steps} & \textbf{Primary Bottleneck Solved}\\
        \hline
        Baseline-1(BL1)   & Direct evaluation of $acos/cos$.  & (N/A - Slowest baseline) \\ \hline
        Baseline-2(BL2)   &PyTorch + cuBLAS \cite{nvidia_cublas_2025}. & (N/A - Strong baseline) \\ \hline
        V1    & Chebyshev recurrence ($T_{n}=2xT_{n-1}-T_{n-2}$)  & Replaces $acos/cos$ calls. \\ \hline
        V2    & V1 + \textbf{Opt 1: LUT Interpolation.} & Eliminates recurrence. \\ \hline
        V3    & V2 + \textbf{Opt 2: Output-Aligned 2D Tiling.} & Mitigates 32-way LUT scatter (reduces to 8-way). \\ \hline
        V4    & V3 + \textbf{Opt 3: Two-Stage Reduction.} & Solves forward-pass atomic contention. \\ \hline
        V5 (PolyKAN)   & V4 + \textbf{Opt 4: Layout Reordering.} & Solves 8-way Coeff stride (enables coalescing). \\   \hline  
    \end{tabularx}
    \label{table_method}
\end{table*}

Baseline-1 evaluates Chebyshev basis via the \allowbreak trigonometric form. 
Baseline-2 (cuBLAS) \cite{ChebyKAN_SynodicMonth_2024} constructs Chebyshev basis 
via the recurrence form and employs the stock PyTorch implementation, which internally dispatches dense GEMM calls to NVIDIA’s cuBLAS library. CuBLAS is widely regarded as the industry-standard, 
vendor-tuned kernel for general-purpose matrix multiplication on GPUs. 
Specifically, PyTorch 2.2 decomposes the layer into a Triton kernel \cite{tillet2019triton} that materializes Chebyshev basis and a GEMM core executed by cuBLAS \cite{nvidia_cublas_2025}. For small tensors Triton fuses both stages into one kernel.
Baseline2 represents the most efficient implementation currently used by ChebyKAN. 
Versions V0 $\rightarrow$ V5 are evaluated cumulatively. \textbf{Each version inherits all preceding optimizations.} 

\subsection{End-to-End Evaluation}
\begin{figure}[htbp]
  \centering
  \includegraphics[width=0.95\linewidth]{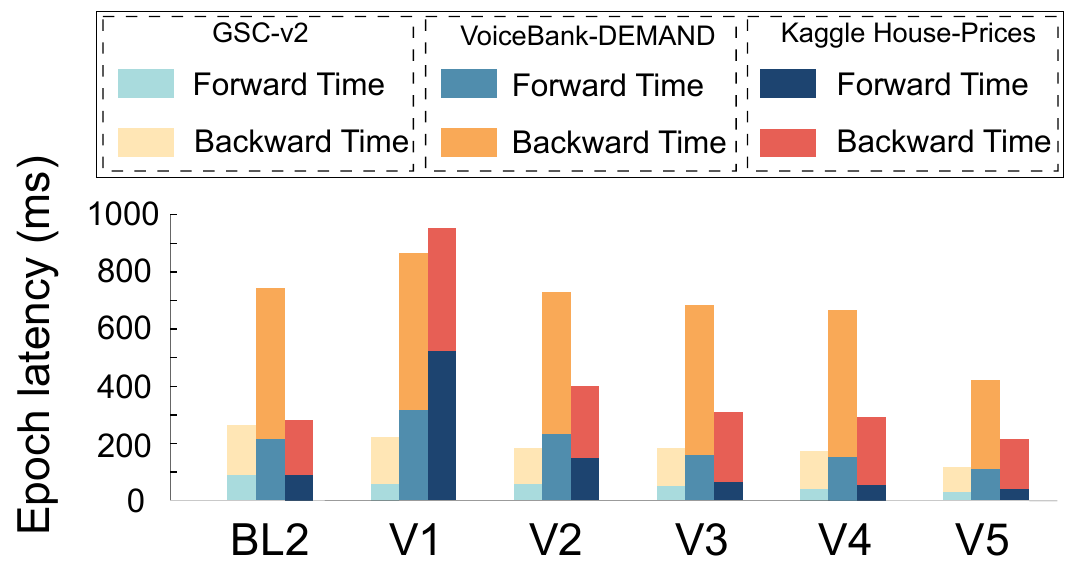}
  \caption{Epoch latency of different versions on three benchmarks.}
  \label{fig_duan}
\end{figure}

To quantify the user-perceived benefit of the proposed optimizations, we measure the end-to-end training and inference cost of all seven kernel 
versions on three workloads. As the performance of Baseline-1 (BL1) is orders of magnitude worse than the other kernel versions, including it in Figure \ref{fig_duan} would severely distort the y-axis and make the results of the other versions difficult to discern. Therefore, we report the results for BL1 separately. The total single-epoch latency for BL1 on the Speech Commands, VoiceBank, and House-Prices tasks are 1798ms, 7193ms, and 3127ms, respectively. Figure~\ref{fig_duan} consolidates, for each workload, the wall-clock time of a single epoch for the remaining kernel versions into two components: forward propagation and backward propagation on the training split.

From Baseline-1 to Baseline-2, replacing the conventional CUDA kernels with a cuBLAS‐based reference 
already reduces epoch time by $7$--$11\times$. Our optimization version V5 shortens the epoch by 
$1.3$--$2.2\times$ compared with Baseline-2.

As detailed in Table \ref{tab:throughput}, V5 demonstrates significant throughput improvements across 
all tasks. For the Google Speech Commands v2 workload, V5 achieves an overall speed-up of $14.8\times$ relative to Baseline-1 and 
$3.9\times$ over Baseline-2. For the VoiceBank-DEMAND dataset, V5 outperforms Baseline-1 and Baseline-2 by factors of
$15\times$ and $2.1\times$, respectively. On the House-Prices regression task, V5 yields a $12.6\times$ gain compared to the initial implementation and a 
$1.3\times$ improvement over the cuBLAS-optimized baseline.

\subsection{Numerical Fidelity and Convergence Speed Evaluation}
A primary concern with replacing an exact mathematical formulation (i.e., Chebyshev recurrence) with an \texttt{LUT}-based interpolation is the potential loss of numerical fidelity. In our end-to-end evaluation, we therefore explicitly compared the final model accuracy of our optimized \textbf{V5 (PolyKAN)} operator against the \textbf{Baseline-2 (BL2)} implementation and MLP implementation. We conducted this analysis on the Speech Commands and Kaggle House Prices tasks. We omitted the VoiceBank-DEMAND task from this comparison, as its simplicity led all models to achieve $100\%$ accuracy within the first few epochs, offering no meaningful differentiation. 

\begin{figure}[h!]
    \centering
    \begin{subfigure}{0.48\textwidth}
        \centering
        \includegraphics[width=\linewidth]{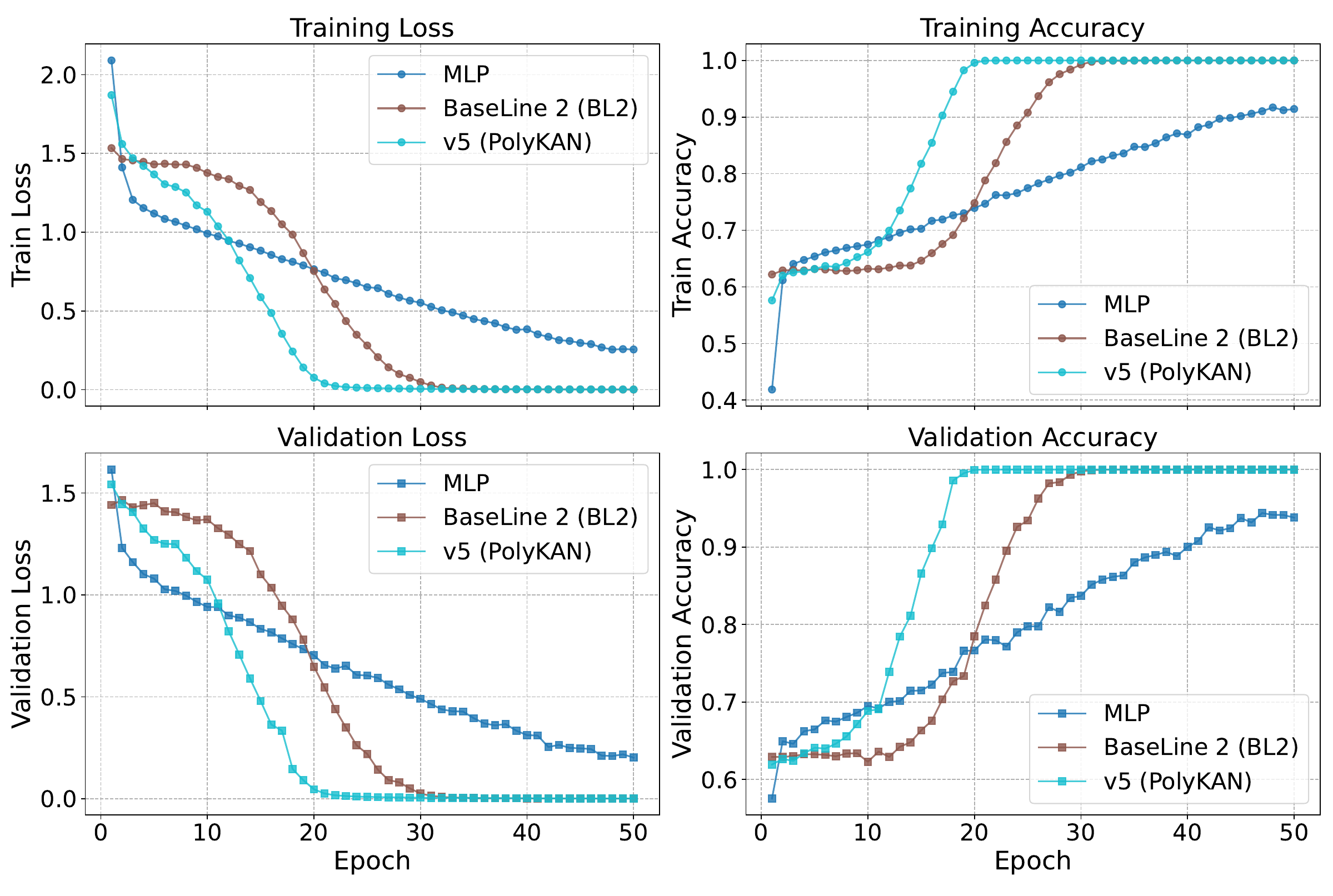}
        \caption{Training and validation loss/accuracy on the GSC-v2 task.}
        \label{fig:gsc_accuracy}
    \end{subfigure}
    
    \vspace{1em}
    
    \begin{subfigure}{0.48\textwidth}
        \centering
        \includegraphics[width=\linewidth]{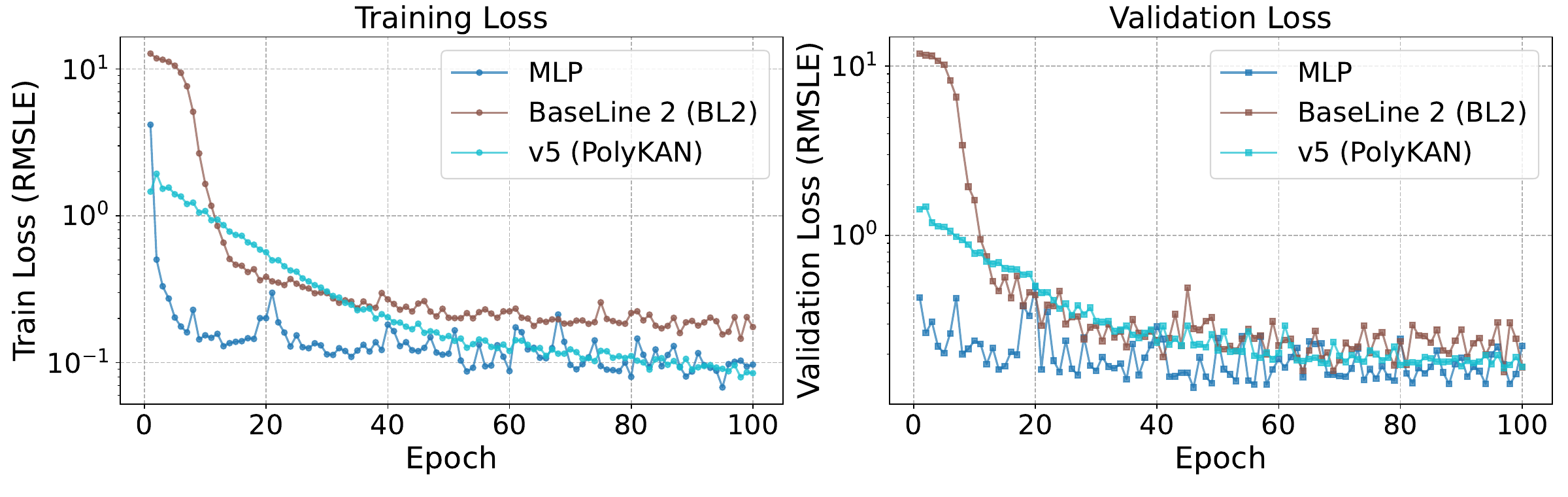}
        \caption{Training and validation loss (RMSLE) on the Kaggle House Price task.}
        \label{fig:house_accuracy}
    \end{subfigure}

    \caption{Comparison of convergence behavior of V5 (PolyKAN) and the recurrence-based BL2 on (a) the GSC-v2 task and (b) the Kaggle House Price task.}
    \label{fig:gsc_house_accuracy}
\end{figure}

The results, presented in Figure \ref{fig:gsc_accuracy} and \ref{fig:house_accuracy}, confirm that our interpolation-based optimizations preserve numerical fidelity. On the House Prices task, PolyKAN achieves a final validation RMSLE comparable to or better than both MLP and BL2. On Speech Commands task, PolyKAN not only matches but ultimately outperforms the BL2 baseline, achieving a higher convergence rate. 

We attribute this accelerated convergence to the fundamental difference in gradient computation. Baseline-2 (Recurrence) uses PyTorch's \texttt{autograd} on the exact recurrence formula , which produces a complex and high-frequency ``jagged'' gradient landscape for high-degree polynomials. Conversely, our V5 (PolyKAN) kernel computes an \textit{approximate, piecewise-constant} gradient derived from the finite difference between \texttt{LUT} sample points (\texttt{(tR - tL) / step}). This ``smoother'' gradient acts as an implicit regularizer, allowing the Adam optimizer to find a more stable and rapid descent path. Our fused operator thus provides a dual benefit: not only is each training step significantly faster, but fewer steps are required to reach the optimal model accuracy.

\begin{table}[htbp]
  \centering
  \caption{Throughput (samples/s) across three kernels.}
  \begin{tabular}{lccc}
    \toprule
    \textbf{Workload} & \textbf{Baseline-1} & \textbf{Baseline-2} & \textbf{V5} \\
    \midrule
    Speech Commands & 5538  & 21029 & 82030 \\
    VoiceBank       & 1451 & 10281 & 21850 \\
    House-Price     &   396   &  3721 &  5000 \\
    \bottomrule
  \end{tabular}
  \label{tab:throughput}

\end{table}

\subsection{Operator-Level Performance Analysis}

Table \ref{tab:layer_latency} contrasts the layer latency of the seven implementations on A100 GPU. We benchmark three precisely defined tensor configurations so that the micro-benchmarks mirror the shapes encountered in the end-to-end experiments:
\begin{itemize}
  \item \textbf{Config. 1}: a mini-batch of 128 samples with 40 features transformed to 256 features, degree 8;
  \item \textbf{Config. 2}: a mini-batch of 64 samples with 256 features transformed to 512 features, degree 15;
  \item \textbf{Config. 3}: a mini-batch of 32 samples with 512 features transformed to 1024 features, degree 24.
\end{itemize}

These three shapes span the three regimes of our workloads, ensuring that layer-level measurements are directly comparable to the end-to-end results.



\begin{table*}[tb]
\centering
\caption{
  Operator-level latency \for a single layer's forward and backward pass, measured on an NVIDIA A100 GPU.
}
\label{tab:layer_latency}
\resizebox{\textwidth}{!}{
\begin{tabular}{l rrr rrr rrr}
\toprule
& \multicolumn{3}{c}{\textbf{Config. 1 (Speech-like)}} & \multicolumn{3}{c}{\textbf{Config. 2 (VoiceBank-like)}} & \multicolumn{3}{c}{\textbf{Config. 3 (HousePrice-like)}} \\
\cmidrule(lr){2-4} \cmidrule(lr){5-7} \cmidrule(lr){8-10}
\textbf{Kernel Version} & \textbf{Fwd (ms)} & \textbf{Bwd (ms)} & \textbf{Sum (ms)} & \textbf{Fwd (ms)} & \textbf{Bwd (ms)} & \textbf{Sum (ms)} & \textbf{Fwd (ms)} & \textbf{Bwd (ms)} & \textbf{Sum (ms)} \\
\midrule
Config. (B, D$_{\text{in}}$, D$_{\text{out}}$, d) & \multicolumn{3}{c}{(128, 40, 256, 8)} & \multicolumn{3}{c}{(64, 256, 512, 15)} & \multicolumn{3}{c}{(32, 512, 1024, 24)} \\
\midrule
Baseline-1 (BL1) & 1.301 & 2.577 & 3.878 & 13.881 & 22.359 & 36.240 & 33.541 & 66.806 & 100.347 \\
Baseline-2 (BL2) & 0.526 & 1.588 & 2.114 & 1.180 & 4.186 & 5.366 & 2.032 & 6.389 & 8.421 \\
V1  & 0.181 & 1.026 & 1.207 & 3.683 & 4.378 & 8.061 & 29.108 & 19.031 & 48.139 \\
V2  & 0.185 & 0.414 & 0.599 & 1.659 & 3.003 & 4.662 & 4.666 & 10.971 & 15.637 \\
V3  & 0.064 & 0.450 & 0.514 & 0.791 & 3.311 & 4.102 & 1.985 & 9.695 & 11.680 \\
V4  & 0.053 & 0.450 & 0.503 & 0.652 & 3.179 & 3.831 & 1.706 & 9.699 & 11.405 \\
\textbf{V5 (PolyKAN)} & \textbf{0.046} & \textbf{0.123} & \textbf{0.169} & \textbf{0.376} & \textbf{1.006} & \textbf{1.382} & \textbf{1.580} & \textbf{4.500} & \textbf{6.080} \\
\bottomrule
\end{tabular}
}
\end{table*}

\begin{figure}[htbp]
  \centering
  \includegraphics[width=0.9\linewidth]{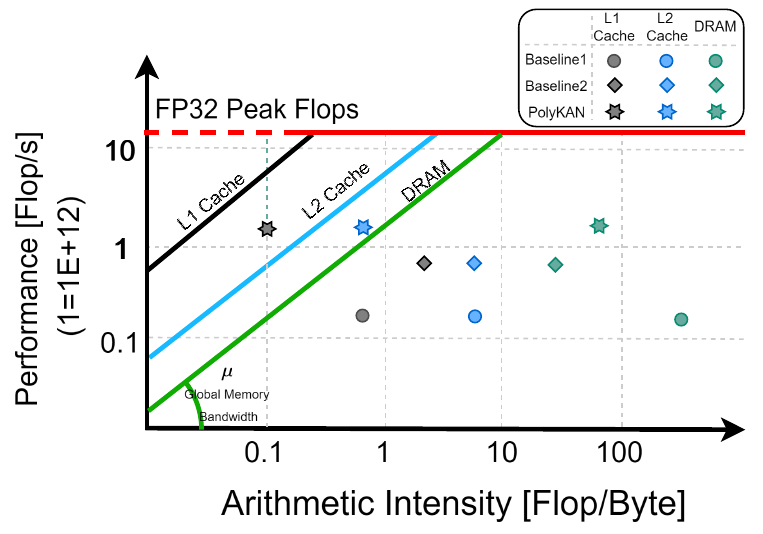}
  \caption{Multi‑Level Roofline Analysis (L1, L2, DRAM) of Baseline Implementations vs. PolyKAN.}
  \label{fig_roofline}
\end{figure}

V5 (PolyKAN) achieves its most dramatic speedup in the small-scale configuration (Config. 1), delivering a $12.5\times$ reduction in total latency over BL2. As problem size and computational density increase (Config. 2 and 3), V5 maintains a robust $1.4$--$3.9\times$speedup over BL2.
Figure \ref{fig_roofline} maps representative kernels onto the roofline model, revealing a coherent migration of the performance bottleneck that validates our strategy. 
The multi\allowbreak‑level roofline plot confirms that PolyKAN attains a superior balance of arithmetic intensity and hardware utilization. By tailoring thread-level parallelism to each stage’s resource mix, PolyKAN sustains consistently high GPU activity.


\subsection{General Applicability Evaluation}
To assess generalization, we apply our acceleration techniques to FourierKAN and benchmark it against the state-of-the-art FusedFourierKAN \cite{FusedFourierKAN2024} accelerator. On an A100 GPU, we measured throughput and end-to-end latency on the Speech Commands task, with results presented in Table \ref{tab:perf_comparison_transposed}.

\begin{table}[tbp]
\centering
\caption{Performance Comparison: FusedFourierKAN vs. OurFourierKAN on Speech Commands Dataset.}
\label{tab:perf_comparison_transposed}

\renewcommand{\arraystretch}{1.1}

\resizebox{\columnwidth}{!}{
\begin{tabular}{lccc}
\toprule
\textbf{Model} & {\shortstack{Forward Latency\\(ms/batch)}} & {\shortstack{Backward Latency\\(ms/batch)}} & {\shortstack{Throughput\\(samples/s)}} \\
\midrule
FusedFourierKAN & 2.76 & 9.63 & 10327 \\
OurFourierKAN   & 0.52 & 1.84 & 54222 \\
\bottomrule
\end{tabular}}
\end{table}


Relative to FusedFourierKAN, our general optimization methods achieved a $5\times$ increase in throughput while simultaneously reducing latency by a factor of 5. In addition to ChebyKAN and FourierKAN, applying the same optimization methodology to other KAN variants, such as LegendreKAN, HermiteKAN, also delivers substantial performance improvements. These results not only substantiate the effect of the proposed optimization paradigm but also demonstrate its portability and general applicability across KAN variants.

\subsection{Discussion}
To demonstrate portability across heterogeneous hardware platforms, we evaluate \textit{ChebyKAN} on a consumer-grade RTX\,4060 GPU. 
The results mirror the trend observed on A100, with our optimized kernel reducing latencies by $1.7$--$2.3\times$ across all configurations.
\section{Conclusion}
We propose an operator-level optimization design for accelerating polynomial KAN variants on GPUs. 
Our fused-kernel approach eliminates computational and memory bottlenecks by integrating four techniques: LUT-based basis evaluation, 2D tiling, a two-stage reduction, and coefficient reordering. Compared to the Python-based implementation and the 
vendor-optimized baseline version, our method significantly improves speed and throughput across three heterogeneous benchmarks. 
The design generalizes broadly to architectures like ChebyKAN and other polynomial architectures, serving as a reusable component for basis function layers.


\bibliography{sample-base}

\end{document}